\documentclass[showpacs,preprintnumbers,amsmath,amssymb]{revtex4}
\usepackage{graphicx}% Include figure files
\usepackage{dcolumn}% Align table columns on decimal point
\usepackage{bm}% bold math
\begin{document}
\title{Unparticle in (1+1) dimension with one loop correction}
% Force line breaks with \\
\author{Anisur Rahaman}
\email{anisur.rahman@saha.ac.in} \affiliation{Hooghly Mohsin
College, Chinsurah, Hooghly - 712101, West Bengal, India}

\date{\today}% It is always \today, today,
             %  but any date may be explicitly specified

\begin{abstract}
The possible emergence of unparticle has been mooted recently
including a mass like term for gauge field with the Schwinger
model at the classical level. A one loop correction due to
bosonization is taken into account and investigation is carried
out to study its effect on the unparticle scenario. It is observed
that the physical mass, viz., unparticle scale  acquires a new
definition, i.e., the effect of this correction enters into the
unparticle scale in a significant manner. The fermionic propagator
is calculated which also agrees with the new scale. It has also
been noticed that a novel restoration of the lost gauge invariance
reappears when the ambiguity parameter related to the current
anomaly acquires a specific expression. We have also observed that
a quantum effect can  nullify the effect of violation of gauge
symmetry caused by some classical terms.
\end{abstract}
%\pacs{11.15Tk, 04.60Kz}
 \maketitle

\section{\bf{Introduction}}
 QED, viz., Schwinger model in $(1+1)$ dimension \cite{SCH}, is
  an interesting and well known
field theoretical model. It has been widely studied over the years
 in connection with the mass generation of
photon, confinement aspect of fermion (quark), charge shielding
etc. \cite{LO, COL, CAS, AG}.  The description of this model in
noncommutative space time \cite{APR, APR1} has also attracted a
considerable attention \cite{SADO, FENG}. The model in a dilatonic
back ground has used to study the information loss problem too
\cite{STORM, AR2}. Though this simple model describes the
interaction of massless fermion with the Abelian gauge field the
physics within it is much involved. Photon acquires mass via a
kind of dynamical symmetry breaking and the fermions gets
disappeared from the physical spectra. The possibility of
emergence of unparticle in (1+1) dimension \cite{GEOR1} has added
a new dimension to the fame of this model. The  interest on this
model concerning the possibility of presence of unpaticle
\cite{GEOR1} has thus opened up further scope of investigation.

 In
general, unparticle is a scale invariant sector, that decouples at
large scale and the spectrum of which might be detected in the
missing energy momentum distribution \cite{GEOR}. Recently in
\cite{GEOR1}, we  have witnessed an exciting illustration where an
attempt was made to establish the possibility of appearence of in
(1+1) dimension and it has been made possible just by the
inclusion of a masslike term for the gauge field into the action
of the Schwinger model at the classical level \cite{GEOR1}. It has
been shown there that it approaches to a free field theory at high
energies and a scale invariant theory at low energy. Despite the
inclusion of the mass term at the classical level there lies more
involved mechanism for a masslike term to get involved at the
quantum mechanical level through a one loop correction \cite{AR}.
An ambiguity parameter automatically enters into that loop
correction. So the effective action of the Schwinger model in its
generalized bosonized version contains a one loop correction along
with an ambiguity parameter \cite{AR, AR1}. For a specific choice,
i.e., for the vanishing value of the ambiguity parameter the model
reduces to the usual gauge invariant vector Schwinger model, but
for the other admissible value of this parameter, the phase space
structure as well as  the the physical spectra acquire significant
changes \cite{AR}. For instance, in this generic situation  the
confinement scenario of the fermion is found to get altered
\cite{AR}. In fact, the fermion gets liberated in the similar
fashion as it was found to happen in the Jackiw-Rajaraman version
of  Chiral Schwinger model \cite{JR}.  We were familiar with this
type of correction along with an ambiguity parameter from long
back \cite{JR}. In \cite{JR, RABIN, GIR, GIR1, GIR2} we noticed,
how amazingly ambiguity parameter removed the long suffering of
the chiral generation of the Schwinger model\cite{HAG}. The
investigation related to unparticle in (1+1) dimension taking
 into account quantum correction would, therefore, be of interest in its
 own right. Inclusion of
 that a quantum correction in QED was attempted by us in
 \cite{AR}. Here we are embarked on a program
  to investigate what new light this one loop effect can
shed in lower dimensional unparticle scenario.

 In (1+1)
dimension, the standard bosonization technique provides a room to
study this newly conjectured unparticle scenario where  both the
classical and one loop corrected masslike term can simultaneously
be taken into account in the same footing. What new signal this
one loop corrected term can convey to this recently proposed lower
dimensional unparticle physics, that is the main objective of this
work. Other perspective of this work is to exploit arbitrariness
of the ambiguity parameter which would be available here due to
this very correction to restore the symmetry of this gauge variant
version which was broken at the classical level  to modify the
standard QED model in order to establish the possible emergence of
unpractice \cite{GEOR1}.

The  model with this setting would, therefore, be the Sommerfield
(Thirring-Wess) \cite{SOMM, THIR} model along with a one loop
correction term. An  illustration with only loop correction was
made by us for a different perspective in \cite{AR}. In this
context, we would like to mention that the masslike term for gauge
gauge field found in the Sommerfield(Thirring-Wess) model
\cite{SOMM, THIR} was included at the classical level. One may
consider it as model that describe the interaction of fermionic
field with Proca background. So there lies a sharp difference
between the mass like term that entered in \cite{AR, AR1} and
\cite{GEOR1, SOMM, THIR} so far  origin of this type of term is
concerned. The masslike term considered in \cite{AR, AR1} comes as
a one loop correction in order to remove the divergence of the
fermionic determinant during bosonization but in the illustration
of unparticle in \cite{GEOR1} the  authors considered the masslike
term for gauge field at the classical level. In fact, the old
model available in \cite{SOMM, THIR} has been considered in
\cite{GEOR1} with a different perspective.

The paper is organized in the following manner. Sec. II is devoted
with the calculation of fermionic current where we will be able to
see how the ambiguity parameter gets involved . In Sec. III, the
bosonized model is quantized using the formalism developed by
Dirac for constrained system to find out the physical spectrum.
Fermionic propagator is calculated in Sec. IV. to see whether it
agrees with the spectrum calculated in Sec. III. In Sec. V, it is
shown that a novel gauge invariance can be restored exploiting the
regularization ambiguity. Sec. VI contains the concluding remark.
\section{{\bf Calculation of fermionic Current}}
The vector Schwinger model is described by the following
generating functional
\begin{equation} Z[A] = \int
d\psi d\bar\psi e^{\int d^2x{\cal L}_f},
\end{equation} with
\begin{equation}
{\cal L}_f = \bar\psi\gamma^\mu[i\partial_\mu + e A_\mu]\psi.
\label{FL}
\end{equation}
The ferminic current for this model is $J_\mu=
\bar\psi\gamma_\mu\psi$. It is a composite operator build out of
fermion fields.  Since the products of local operators are often
singular, it is instructive to calculate the current by point
splitting method.  Maintaining the Lorentz invariance the current
can be defined in the following manner.
\begin{equation}
J_\mu =\bar{\psi}(x+\epsilon)\gamma_\mu:e^{ie\int_{x}^{x+\epsilon}
dz^\mu[(1+a)A_\mu(z) - a\tilde A_\mu(z)]}:\psi(x), \label{DCC}
\end{equation}
where $\tilde A_\mu =\epsilon_{\mu\nu} A^\nu$. The choices of
coefficients in front of $A_\mu$ and $\tilde A_\mu$ are very
crucial. Note that $a=0$, correspond to the usual gauge invariant
vector Schwinger model, but the non zero $a$'s, correspond to the
nonconfining Schwinger model proposed by us in \cite{AR}.  In
(1+1)dimension, the most general Ansatz for $A_\mu$ is
\begin{equation}
A^\mu= -\frac{\sqrt\pi}{e}(\tilde\partial_\nu\sigma +
\partial_\nu\tilde\eta), \label{DA}
\end{equation}
where $\tilde\partial_\mu = \epsilon_{\mu\nu}\partial^\nu$.
$\sigma$ and $\eta$ are two scalar field and the dual of $\eta$ is
defined by $\tilde\partial_{\mu} \eta = \partial_{\mu}\tilde\eta$.
The definition (\ref{DA}) of $A_\mu$   gives
\begin{equation}
F_{\mu\nu}=-\frac{\sqrt\pi}{e}(\partial_\mu\tilde\partial_\nu -
\tilde\partial_\mu\partial_\nu)\sigma =
\frac{\sqrt\pi}{e}\epsilon_{\mu\nu}\Box\sigma,
\end{equation}
The equation of motion, that follows from (\ref{FL}), for the
field $\psi$ is
\begin{equation}
[i\partial\!\!\!/ + \sqrt{\pi}\tilde\gamma_\mu(\sigma +\eta)]\psi
=0,
\end{equation}
and it provides the following solution for $\psi$
\begin{equation}
\psi(x)=:e^{i\sqrt{\pi}\gamma_5(\sigma(x)+\eta(x))}:\psi^{(0)}(x).
\label{DPSI}
\end{equation}
If we substitute the expression for $A_\mu$ and  $\psi$  in
(\ref{DCC}), and use the identity
\begin{equation}
<0|{\bar\psi}_\alpha^{(0)}(x+\epsilon)\psi_\beta^{(0)}(x)|0> =
-i\frac{(\epsilon^\mu\gamma_\mu)_{\alpha\beta}} {2\pi\epsilon^2},
\end{equation}
we  arrive at
\begin{equation}
J_\mu^{reg} = J_\mu^f -
\frac{a}{\sqrt\pi}[\frac{\epsilon_\mu\epsilon_\nu -
\tilde\epsilon_\mu\tilde\epsilon_\nu}
{\epsilon^2}\tilde\partial_\nu(\sigma +\eta)].
\end{equation}
Taking the symmetric limit, i.e., averaging over the point
splitting direction $\epsilon$, we find
\begin{equation}
J_\mu^{reg}=J_\mu^f -\frac{a}{\sqrt{\pi}}\tilde\partial_\mu(\sigma
+ \eta))\label{RFC}
\end{equation}
In terms of potential of the fermionic current, the above equation
can be written down as
\begin{equation}
J_\mu^{reg}= - \tilde\partial_\mu \phi +
\frac{ea}{\sqrt{\pi}}A_\mu.\label{RCC}
\end{equation}
Here $\phi$ stands for the potential of the free fermionic current
and $ J_\mu^f = :{\bar\psi}^{(0)}(x)\gamma_\mu\psi^{(0)}$.
 Note
that the expression of current (\ref{RCC})contains the ambiguity
parameter $a$. We were familiar with this parameter from long past
\cite{JR}. In \cite{AR, AR1} we have argued that this type of
ambiguity may appear in the Schwinger model too however specific
calculation was lacking there. An arguments illustration as given
there like the argument made in \cite{JR}. For the the second case
\cite{JR} it was calculated in \cite{RABIN} after few years. Rest
of the work in this paper is devoted  along with the ambiguity
parameter $a$ to study the unpractice scenario in (1+1) dimension.

\section{{\bf Study of physical mass through constraint analysis}}
 In (1+1) dimension exact Bosonization is possible and it interesting
 that one loop effect automatically enters through the process and
 that enable us to study the effect of one loop correction in the lower
  dimensional unparticle scenario. So we start our analysis
  with a generalized bosonized version where corrections from both the corners
  (classical as well as quantum) are taken into consideration. A
  standard way
  to get the  bosonized
  lagrangian density from
(\ref{FL}), is to be integrating out the fermions one by one
\cite{AR}, and that leads to the following bosonized lagrangian
density
\begin{equation}
{\cal L}_B = \frac{1}{2}\partial_\mu\phi \partial^\mu\phi -
e\epsilon_{\mu\nu}\partial^\nu\phi A^\mu + \frac{1}{2}ae^2
A_{\mu}A^{\mu}. \label{BLD}
\end{equation}
Here $a$ is the regularization ambiguity parameter. It is needed
to remove the divergence in the fermionic determinant. When it is
treated through appropriate regularization, a masslike counter
term enters in to the model through a one loop correction
\cite{AR}. The current for this model is found out to be
\begin{equation}
J_\mu^{B}= - e\tilde\partial_\mu \phi + e^2aA_\mu.
\end{equation}
 Note that this expression of current
 agrees with the current computed in the previous section
(\ref{RCC}), and it is straight forward to see that $\partial_\nu
J_\mu^{B} \ne 0$ that indicates the violation of  gauge symmetry
at the bosonized action due to the one loop effect. However at the
fermionic level current was $J_\mu= \bar\psi\gamma_\mu\psi$ and
gauge symmetry was intact there from classical point of view.

  If Proca type  electromagnetic background
 (along with the corresponding masslike term)  for the
gauge field is added with the fermionic version of the Schwinger
model the model turns into
\begin{equation}
{\cal L}_{fg} = \bar\psi\gamma^\mu[i\partial_\mu + e A_\mu]\psi +
\frac{1}{2} m_0 A_{\mu}A^{\mu}.
\end{equation}
and gauge symmetry immediately breaks here at the classical level
due to the presence masslike term of Proca background. As we have
already mentioned ,   the standard bosonization technique allows
us to take the advantage of including a one loop (quantum) effect
here too. If we now include the one loop correction the bosonised
lagrangian density reads.
\begin{equation}
{\cal L}_B = \frac{1}{2}\partial_\mu\phi \partial^\mu\phi -
e\epsilon_{\mu\nu}\partial^\nu\phi A^\mu -
\frac{1}{4}F_{\mu\nu}F^{\mu\nu} + \frac{1}{2}e^2(a +
\frac{m_0}{e^2}) A_{\mu}A^{\mu}. \label{BLM}
\end{equation}
Here Lorentz indices runs over the two values $0$ and $1$
corresponding to the two space time dimension and the rest of the
notations are standard. The antisymmetric tensor is defined with
the convention $\epsilon_{01}=+1$. The coupling constant $e$ has
one mass dimension in this situation. The parameter $m_0$ of
course, has the dimension $e^2$.  The terms containing  $m_0$, has
been introduced here to get the possible presence the unparticle
is it was done in \cite{GEOR1}. Of course it was considered
initially in the Sommerfield (Thirring-Wess) model \cite{SOMM,
THIR}. The term containing $a$ however help to see the possible
change of the unparticle scenario.  Note that gauge symmetry has
been broken by the combined effect of the parameter $a$ and $m_0$.

Let us now proceed to study the phase space structure of the model
in order to observe how physical mass term of the model
\cite{GEOR} changes with the presence of one loop corrected
masslike term. Though it can be seen in various ways here we will
follow the standard quantization of constrained system developed
by Dirac \cite{DIR}. To this end, it is necessary to calculate the
the momenta corresponding to the field $A_0$, $A_1$ and $\phi$.
From the standard definition of the momentum, we obtain
\begin{equation}
\pi_0=0 \label{M1},
\end{equation}
\begin{equation}
\pi_1 = F_{01}\label{M2},
\end{equation}
\begin{equation}
\pi_\phi = \dot\phi - eA_1\label{M3},
\end{equation}
where $\pi_0$, $\pi_1$ and $\pi_\phi$ are the momenta
corresponding to the field $A_0$, $A_1$ and $\phi$ respectively.
Using the equations (\ref{M1}), (\ref{M2}) and (\ref{M3}), the
canonical hamiltonian density is calculated:
\begin{equation}
{\cal H}_c = \frac{1}{ 2}(\pi_\phi +eA_1)^2 + \frac{1}{2}\pi_1^2 +
\frac{1}{2}\phi'^2 + \pi_1A_0' - eA_0\phi' - \frac{1}{2}e^2(a +
\frac{m_0}{e^2})(A_0^2 - A_1^2).\label{CHAM}
\end{equation}
Note that $\omega = \pi_0 \approx 0$, is the familiar primary
constraint of the theory. The preservation of the constraint
$\omega(x)$ requires $[\omega(x), H(y)] = 0$, which leads to the
Gauss' law as a secondary constraint:
\begin{equation}
\tilde\omega = \pi_1' + e\phi' + e^2(a + \frac{m_0}{e^2})A_1
\approx 0. \label{SCO}
\end{equation}
The constraints (\ref{M1}) and (\ref{SCO}), form a second class
set. Treating (\ref{M1}) and (\ref{SCO}), as strong condition one
can eliminate $A_0$, and obtain the reduced hamiltonian density.
\begin{equation}
{\cal H}_r = \frac{1}{2}(\pi_\phi + eA_1)^2 +
\frac{1}{2m_0}(\pi'_1 + e\phi')^2 + \frac{1}{2}({\pi_1}^2 +
\phi'^2)^2 + \frac{1}{2}e^2(a + \frac{m_0}{e^2})A_1^2.
\label{RHAM}
\end{equation}
According to the Dirac's prescription  of quantization \cite{DIR},
the Poisson brackets become inadequate for a theory possessing
second class constraint in its phase space. This type of system
however, remains consistent with the Dirac brackets \cite{DIR}. It
is straightforward to show that the Dirac brackets between the
fields describing the reduced hamiltonian (\ref{RHAM}) remain
canonical. Using the Dirac brackets,  the following first order
equations of motion  are found out from the reduced Hamiltonian
density (\ref{RHAM}).
\begin{equation}
\dot A_1= \pi_1 -\frac{1}{e^2(a + \frac{m_0}{e^2})}(\pi_1'' +
e\phi''), \label{EQM1}
\end{equation}
\begin{equation}
\dot\phi = \pi_\phi + eA_1,\label{EQM2}\end{equation}
\begin{equation}
\dot \pi_\phi = (1+ \frac{1}{(a + \frac{m_0}{e^2})})\phi'' +
\frac{1}{e(a + \frac{m_0}{e^2})}\pi_1'' , \label{EQM3}
\end{equation}
\begin{equation}
\dot\pi_1 = -e\pi_\phi - e^2(a + \frac{m_0}{e^2}+1)A_1.
\label{EQM4}
\end{equation}
 A little algebra converts the
above first order equations (\ref{EQM1}), (\ref{EQM2}),
(\ref{EQM3}) and (\ref{EQM4}), into the following two second order
differential equations:
\begin{equation}
[\Box + e^2(a + \frac{m_0}{e^2}+ 1)]\pi_1 = 0, \label{SP1}
\end{equation}
\begin{equation}
\Box[\pi_1 + e(a + \frac{m_0}{e^2}+ 1)\phi] = 0. \label{SP2}
\end{equation}
Equation (\ref{SP1}), describes a massive boson field with square
of the  mass $m^2 =e^2(a+\frac{m_0}{e^2}+1)$, whereas,  equation
(\ref{SP2}), describes a massless scalar field. The equation
(\ref{SP1}), clearly shows that the physical mass acquires a
generalized expression with the parameters involved in the
masslike term at the classical level, as well as with the
parameter entered through the one loop correction during
bosonization. The  welcomed entry of the  one loop correction has
thus brought a noticeable change in the physical mass and
consequently, the definition of unparticle scale \cite{GEOR1}
acquires a novel expression with an adjustable parameter $a$.

Regarding the appearance of $a$ dependent term,  we should make
some comment in order to the make fact transparent on one side,
and to avoid confusion on the other. In two dimensional field
theory, it is very crucial to define fermionic current, because
lots of subtleties are involved in it. With the definition of the
fermionic current, the mathematical structure of the effective
action changes significantly. In \cite{ADAS}, it was  pointed out
that there was no arbitrariness in the chiral Schwinger model and
that result corresponds to $a=0$ of the Jackiw-Rajaraman version
of chiral Schwinger model where it was maintained that for $a=0$
the model suffered from non unitrity and that suffering could be
removed only through non vanishing $a$. Few years later, through
appropriate calculation of fermionic current it was  shown in
\cite{RABIN}, that arbitrariness was certainly there and the $a$
dependent counter term resulted in. One can certainly calculate
the fermionic current in various ways for the Schwinger model too.
As one way calculation is exihibited in Sce. II.  For Schwinger
model,  it is an usual practice to , to  defined the point
splitting  maintaining gauge symmetry and a generalized definition
of current maintaining the gauge symmetry is available in
\cite{AG}. However, one can certainly define it ignoring that
symmetry. In the generalized gauge invariant definition \cite{AG}
we  found the emergence of a parameter dependent kinetic energy
like term. The way Hagen defined the current,  did not in any way
include that possibility \cite{HAG1, HAG2}. So all the subtleties
are not included even within the {\it generalised} definition of
Hagen \cite{HAG1, HAG2}. The way we have defined the current in
the present work, is also exclusive of the definition found in
\cite{HAG1, HAG2}, because the definition available in \cite{HAG1,
HAG2} disagrees with non vanishing $a$.

\section{{\bf Calculation of fermionic propagator}}
The nature of the theoretical spectrum becomes more transparent if
we calculate the fermionic propagator to which we will now tern.
To calculate fermion propagator one needs to work with the
original fermionic model. The calculation of fermionic propagator
is analogous to the  Chiral Schwinger model \cite{GIR, GIR1}, and
the so called nonconfining Schwinger model \cite{AR, AR1}. The
effective action obtained by integrating out $\phi$ from the
bosonized action (\ref{BLM}), is
\begin{equation}
S_{eff}=\int d^2x\frac{1}{2}[A_\mu(x)
M^{\mu\nu}A_\nu(x)],\end{equation} where
\begin{equation}
M^{\mu\nu}=  e^2(a + \frac{m_0}{e^2})g^{\mu\nu} - \frac{\Box
+e^2}{\Box}\tilde\partial^\mu\tilde\partial^\nu.\end{equation}
%Here we have used the standard notation
%$\tilde\partial^\mu=\epsilon^{\mu\nu}\partial_\nu.$
 The gauge field propagator is just
the inverse of $M^{\mu\nu}$ and it is found to be
\begin{equation}
\Delta_{\mu\nu}(x-y)=\frac{1}{e^2(a +
\frac{m_0}{e^2})}[g_{\mu\nu}+ \frac{\Box +e^2}{ \Box(\Box + e^2(a
+ \frac{m_0}{e^2}+ 1))} \tilde\partial_\mu\tilde\partial_\nu
]\delta(x-y).
\end{equation}
Note that the two poles of propagator are found at the expected
positions. One is at zero, and another is at $e^2(a +
\frac{m_0}{e^2}+ 1)$, indicating a  massless and a massive
excitations respectively and it agrees with physical spectrum
obtained in the previous Sec.

We are now in a stage to calculate the Green function of the Dirac
operator. To do that, let us consider the following Ansatz  for
the Green function of the Dirac operator $(i\partial\!\!\!\!/ +
eA\!\!\!\!/)$.
\begin{equation}
G(x,y;A)=e^{ie(\Phi(x)-\Phi(y))}S_F(x-y),\label{ANST}
\end{equation} where $S_F$  is  the  free massless fermion
propagator and $\Phi$ will be determined when the Ansatz
(\ref{ANST}) will be plugged into the equation for the Green
function.
 It will enable us to construct the propagator of the
original fermion $\psi$.
 From the
standard construction, the Green function is found out as
\begin{equation}
G(x,y;A)=e^{ie\int d^2zA^\mu(z)J_\mu(z)}S_F(x-y),\end{equation}
where the fermionic {\it current} $J_\mu$ is given by the
following expression.
\begin{equation}
J_\mu=(\partial^z_\mu  +  \gamma_5\tilde\partial^z_\mu)(D_F(z- x)
-D_F(z-y)).\end{equation} Here $D_F$ represents the propagator of
a massless free  scalar  field. In order to avoid singularity such
propagators have to be infra-red regularized in $(1+1)$ dimensions
\cite{LO}.
\begin{equation}
D_F(x)=-\frac{i}{4\pi}ln(-\mu^2x^2+i0).\end{equation} Here $\mu$
stands for the infra-red regulator mass. Finally, we obtain the
fermion  propagator by functionally integrating $G(x,y;A)$ over
the gauge field:
\begin{eqnarray}
S'_F&=&\int {\cal D}A e^{\frac{i}{2}\int d^2z~(A_\mu(z)
M^{\mu\nu}A_\nu(z) + 2eA_\mu   J^\mu)}S_F(x-y)\nonumber\\&=&{\cal
N} \exp[\frac{D_F}{(a + \frac{m_0}{e^2})(a + \frac{m_0}{e^2}+ 1)}+
\frac{\Delta_F(m^2=e^2(a+\frac{m_0}{e^2}+1))}{a + \frac{m_0}{e^2}+
1}]S_F.\label{PROP}
\end{eqnarray} Here $\Delta_F$ is the propagator of a massive free
scalar field and ${\cal N}$ is a wave function renormalization
factor.

The theoretical spectrum (\ref{SP1}) and (\ref{SP2}), as well as
the fermionic propagator (\ref{PROP}), therefore, make the fact
confirm that the mass acquires a generalized expression with bare
coupling constant and the parameters involved within the masslike
terms for the gauge field. Thus, a new definition of unparticle
scale emerges out with the introduction of one loop correction
holding the hands of generalized physical mass term (\ref{SP1}).
It is fascinating to get this new definition of unparticle scale
when one loop correction is taken into consideration and it would
be natural to see its importance in  the area of physics where
unparticle is expected to show its prominent role.

Note that,  the mass, that generates via dynamical symmetry
breaking in the  vector Schwinger model, depend only on the bare
coupling constant, and it does not contain any such parameters
like $m_0$ or $a$ . In fact, in the bosonized vector Schwinger
model, no such parameter gets involved when bosonization is done
there maintaining the gauge symmetry. It is true that setting the
value of the parameter to $a=0$, one can obtain vector Schwinger
model \cite{SCH} from the bosonized lagrangian of the so called
nonconfining Schwinger model \cite{AR}, where bosonization is done
ignoring the gauge symmetry. But, this trivial choice $a=0$, does
not work to bring back the gauge symmetry in this generalized
situation where masslike terms from both the corners classical as
well as quantum are present. However, there lies some interesting
possibility in connection with the retrieval of the gauge symmetry
in the usual phase space, which we are going to uncover in the
following Sec.

\section{{\bf Restoration of gauge symmetry exploiting the ambiguity parameter}}
To exploit arbitrariness of the ambiguity parameter present in a
model is more or less  a standard practice in $(1+1)$ dimensional
field theory. It has been used several times in different
perspective. Sometimes even it has been used as a remedy in order
to get out of the dangerous unphysical situation through which it
is suffering \cite{JR, RABIN, GIR, GIR1, GIR2, PM, MG, KH, ABD}.

To get back the broken gauge symmetry of this modified model, if
we exploit the available arbitrariness of the ambiguity parameter
in this situation it will also stand as a sound example in its own
right. Without violating any physical principal, the said
arbitrariness allows us to  set
\begin{equation}(a +
\frac{m_0}{e^2}) =0. \label{AA}
\end{equation}
This equation  though look simple, it bears a deeper meaning. It
reflects the competition between a classical and quantum
mechanical term. During the competition if the effect one we
succeeds to nullify the effect made by the other then only
equation (\ref{AA}) will be satisfied.  Now what it renders, is
the opening up of a provision of getting an expression
(definition) of $a$ in terms of the parameters involved in the
theory, and that in terns, brings a remarkable change in the
constraint structure of the theory. In fact, this definition of
$a$ corresponds to a singularity in the phase space structure of
the theory. The existing second class constraint structure gets
converted into a first set.  The following are those two first
class constraints if (\ref{AA}) is maintained.
\begin{equation}
\pi_0 \approx 0 \label{FM},
\end{equation}
\begin{equation}
\pi_1' + e\phi' \approx 0. \label{FG}
\end{equation}
So, with this particular expression of the ambiguity parameter
$a$, we get only the two first class constraints (\ref{FM}) and
(\ref{FG}). We now need two gauge fixing conditions to single out
the physical degrees of freedom. The following two conditions
would be the appropriate gauge fixing at this stage.
 \begin{equation}
 A_0 \approx 0, \label{G1}
\end{equation}
 \begin{equation}A_1 \approx 0. \label{G2}
 \end{equation}
When the gauge fixing conditions (\ref{G1}) and (\ref{G2}), are
plugged as a strong condition into the hamiltonian (\ref{CHAM}),
along with the two first class constraints (\ref{FM}) and
(\ref{FG}), the hamiltonian (\ref{CHAM}) acquires the following
simplified form.
\begin{equation}
H_{RF}= \int dx[\frac{1}{2}\pi_{\phi}^2 + \frac{1}{2}e^2\phi^2 +
\frac{1}{2}\phi'^2]. \label{SIMH}
\end{equation}
But the price to pay at this point is to use the Dirac brackets
\cite{DIR} in place of canonical Poisson brackets. The Dirac
brackets \cite{DIR} for the fields describing the reduced
hamiltonian (\ref{SIMH}), are found to remain canonical. We get
the following equations of motion hamiltonian (\ref{SIMH}), when
the canonical Dirac brackets are used for computation of equations
of motion.
\begin{equation}
\dot \phi = \pi_\phi  \label{SEQM3},
\end{equation}
\begin{equation}
\dot \pi_\phi = \phi'' -  e^2\phi. \label{SEQM4}
\end{equation}
The above two equations give a  single Klein-Gordon type second
order differential equation
\begin{equation}
[\Box + e^2]\phi = 0. \label{NSPEC}
\end{equation}
It describes a massive boson with square of the  mass $e^2$. Note
that, the mass in this situation is independent of the parameter
$a$ or $m_0$. The presence of two first class constraint in the
phase space indicates that the model with this  has got back the
gauge symmetry in its usual phase space when the mutual effect of
the classical and quantum mechanical term cancels each other. The
first class constraints will provide the generator of the gauge
transformation. What can be more interesting than to see the
reappearance of the  lost symmetry of a model just by exploiting
the arbitrariness of the ambiguity parameter? A careful look
reveals that the expression of $a$ available by exploiting the
arbitrariness in the theory (\ref{AA}), maps the present model
onto the usual vector Schwinger model.  A trivial choice, i.e.,
simultaneous setting of the parameters $a=0$ and $m_0=0$, of
course, convert this modified model into the usual Vector
Schwinger model, however that choice does not bear any meaningful
physical consequence as it is available for the present context
exploiting the arbitrariness involved in the theory.

 There are different ways to restore gauge invariance of a
 gauge nonsymmetric
model. Mitra-Rajaraman prescription \cite{MR1, MR2} is a good
example of that, where  restoration of gauge symmetry of a given
model is possible in its usual phase space. Some other
prescriptions are also available, where restoration of gauge
symmetry takes place in the extended phase space \cite{FR1, FR2,
FR3, FR4}. However, the restoration of gauge symmetry through a
specific non vanishing expression of the ambiguity parameter $a$
resulted out of the contest of the two same type of terms coming
from two different corner looks amazing.  It is expected that it
would be of considerable importance elsewhere, especially, in the
physics at high energy regime, where unparticle is expected to
show important role.

\section{{\bf Conclusion}}
 Our work reveals that with this new setting the physical mass of the boson field
  acquirs
 a generalized expression. Here physical mass term does not contain only the parameter
 involved in the classical masslke term like \cite{SOMM, THIR,
 GEOR} but also it has acquired the effect of one loop correction. In
 fact, the  physical mass term of the model described
 earlier in
  the work \cite{SOMM, THIR, GEOR1} has got modified with the parameter
 involved in one loop corrected term. So the quantum correction will now
  play a crucial role. Thus a novel definition in the unparticle
 scale and that too with an adjustable parameter $a$ has resulted
 in. So this new setting is expected to encompass both the
 theoretical and phenomenological aspects.

We have  obtained an expression of the ambiguity parameter $a$
  exploiting the arbitrariness  available by virtue of exact
  bosonization
without violating any physical principle which stands as a
fascinating aspect of this model. The service, which this new
definition of $a$   renders is remarkable: it has brought back the
symmetry of the effective action in the usual phase space when
classical masslike term for gauge fields destroyed that symmetry
to start with at the fermionic version. So an opening has resulted
in  where a contest between  a classical term and a quantum
mechanically generated term can be examined in the context of
symmetry restoration or violation. Some more studies that can be
followed from the work are as follows. What would be the fate of
that unparticle for that specific value of $a$ that brings back
the symmetry? Will it die out when the system regain its original
symmetry?

It would be worthwhile to mention that the  term $\frac{1}{2}e^2(a
+ \frac{m_0}{e^2})A^\mu A_\mu$ though looks like a gauge fixing
term, it is not the case in true sence. With this term, the gauge
redundancy of the vector Schwinger model though gets fixed and it
turns into a gauge non invariant model, nevertheless, it can not
be considered as proper gauge fixing because a true gauge fixing
term never brings any change in the theoretical spectrum. in
contrast to that what we noticed here that with the inclusions of
the masslike terms, the mass of the massive boson has acquire a
generalized expression. A massless boson has also been found  in
the theoretical spectrum. Instead of adding the terms
$\frac{1}{2}e^2(a + \frac{m_0}{e^2})A^\mu A_\mu$, if the terms
$\frac{\alpha}{2e^2}(\partial_\mu A^\mu)^2$, are added into the
starting lagrangian of the vector Schwinger model, these do work
as a true gauge fixing term, since in that situation one gets only
a parameter free mass term for the photon, though the starting
lagrangian in that situation contains the parameter $\alpha$.

One more point which we would like to emphasize is that the way
the arbitresses of the ambiguity parameter has been exploited here
to get back the symmetry of the model, has not in any, violated
any physical principle and the technique is more or less standard
in (1+1) dimensional QED and Chiral QED.  We have witnessed
several examples of the use of this mechanism in order to get
rescued from different unfavorable as well as un physical
situations \cite{JR, RABIN, GIR, GIR1, GIR2, PM, MG, KH, ABD}. The
most remarkable one in this context was the Jackiw and Rajaraman
version of chiral Schwinger model \cite{JR}, where they saved the
long suffering of the chiral generation of the Schwinger model
\cite{HAG}, from the non-unitary problem.

\section{\bf{Acknowledgments}}
I would like to thank the Director, Saha Institute of Nuclear
Physics for using computer facilities of the Institute.

\end{document}